# Analysis of Diffractive Neural Networks for Seeing Through Random Diffusers

*Yuhang Li*[1,2,3]        e-mail: yuhangli@g.ucla.edu

*Yi Luo*[1,2,3]           e-mail: yluo2016@g.ucla.edu

*Bijie Bai*[1,2,3]        e-mail: baibijie@g.ucla.edu

*Aydogan Ozcan*[1,2,3]*   e-mail: ozcan@ucla.edu

[1]Electrical and Computer Engineering Department, University of California, Los Angeles, California 90095, USA

[2]Bioengineering Department, University of California, Los Angeles, California 90095, USA

[3]California Nano Systems Institute (CNSI), University of California, Los Angeles, California 90095, USA

*Correspondence: Prof. Aydogan Ozcan

E-mail: ozcan@ucla.edu

Address: 420 Westwood Plaza, Engr. IV 68-119, UCLA, Los Angeles, CA 90095, USA

Tel: +1(310)825-0915

Fax: +1(310)206-4685

# Abstract

Imaging through diffusive media is a challenging problem, where the existing solutions heavily rely on digital computers to reconstruct distorted images. We provide a detailed analysis of a computer-free, all-optical imaging method for seeing through random, unknown phase diffusers using diffractive neural networks, covering different deep learning-based training strategies. By analyzing various diffractive networks designed to image through random diffusers with different correlation lengths, a trade-off between the image reconstruction fidelity and distortion reduction capability of the diffractive network was observed. During its training, random diffusers with a range of correlation lengths were used to improve the diffractive network's generalization performance. Increasing the number of random diffusers used in each epoch reduced the overfitting of the diffractive network's imaging performance to known diffusers. We also demonstrated that the use of additional diffractive layers improved the generalization capability to see through new, random diffusers. Finally, we introduced deliberate misalignments in training to "vaccinate" the network against random layer-to-layer shifts that might arise due to the imperfect assembly of the diffractive networks. These analyses provide a comprehensive guide in designing diffractive networks to see through random diffusers, which might profoundly impact many fields, such as biomedical imaging, atmospheric physics, and autonomous driving.

# Introduction

Imaging through diffusive media is one of the most important and challenging topics in computational imaging[1], [2] with a far-reaching impact in many different applications, such as biomedical imaging [3], atmospheric physics[4], and robotics[5]. If prior knowledge of a diffuser's transmission matrix is accessible[6], [7], deconvolutional algorithms can be used to compute the original image. However, getting an accurate transmission matrix is hard to achieve in many cases[8]. Using guide-stars or reference objects, adaptive optics-based wavefront shaping techniques have been used to image through diffusers[9]. If only a single shot image is available, digital methods utilizing the memory effect of the diffusive medium can also be used to reconstruct an image of the sample[10]–[12]. With the resurgence of deep learning techniques, various new computational methods based on deep neural networks were also implemented [13]–[15], which were trained using random diffusers with a given correlation length to reconstruct the images of objects seen through unknown diffusers of the same correlation length.

All of these methods mentioned above require powerful digital computers to rapidly reconstruct unknown objects behind diffusers. As an alternative approach, we have recently reported an all-optical solution to instantly see through unknown phase diffusers using diffractive deep neural networks ($D^2$NNs), without the need for any digital computer, memory, or external power, except for the illumination light[16]. This $D^2$NN framework is an all-optical machine learning platform that computes the desired task between its input and output fields-of-view (FOVs) using engineered light-matter interaction[17]. The information to be processed is encoded in the phase and/or amplitude of the incident complex optical field of the diffractive network, which is composed of successive transmission and/or reflection surfaces. Each of these diffractive surfaces is composed of, e.g., tens of thousands of independent field modulation units (termed diffractive neurons) designed using deep learning and error back-propagation methods to optimize the amplitude and/or phase profile of the diffractive layers. Together with the free-space propagation in between, these diffractive layers collectively map the input optical field to the desired output profile, which is determined by the task or training loss function. Diffractive networks have been successfully implemented for e.g., object recognition [17]–[19], hologram reconstruction[20], quantitative phase imaging (QPI)[21], single-pixel machine vision[22], performing logic operations[23] and designing a cascadable NAND gate[24]. Extensions of the $D^2NN$ framework to utilize broadband input light were also demonstrated to design non-intuitive optical components for, e.g., spatially-controlled wavelength division multiplexing, spectral filter design[25], and pulse shaping[26].

In our previous work, we experimentally demonstrated the capability to instantly see unknown objects through randomly-generated unknown phase diffusers using diffractive networks operating at the THz part of the spectrum[16]. In each training epoch of one of these $D^2NN$ designs, the images of handwritten digits from the MNIST dataset[27] were distorted by *n* different random diffusers with the same correlation length $L_1$. After being trained for 100 epochs and seeing $N=100n$ different randomly generated phase diffusers, the four-layered diffractive network was trained to map an unknown, random diffuser-distorted optical field onto an output intensity profile that matches the original distortion-free image. Most importantly, the converged network was proven to instantly image through new random diffusers that were never seen in the training process, as long as the correlation lengths of the new diffusers $L_2$ was $\geq L_1$[16]. We also showed that the trained diffractive optical network converged to a general-purpose imager with a strong resilience against random distortions from the diffusers and worked as an imager with and without the presence of diffusers. Compared to traditional learning-based or iterative image reconstruction algorithms implemented on digital computers, a diffractive network does not require computing power except for the illumination light, and completes its image reconstruction task instantly as the input light passes through a thin diffractive volume.

This paper presents a detailed analysis of diffractive neural networks for imaging through diffusers and explores different training strategies and design parameters (see Fig. 1). To better understand the generalization capabilities of a trained diffractive network to image unknown objects through random unknown diffusers, we trained several diffractive networks using diffusers with different correlation lengths $L_1$ and tested them with new, unknown diffusers with a large range of correlation lengths covering both $L_2 < L_1$ and $L_2 \geq L_1$. We demonstrated that the trained diffractive networks could see through new random diffusers with a different statistical distribution from those used in training. This analysis also revealed a trade-off mechanism between the reconstruction fidelity and the distortion reduction capability of trained diffractive networks. This trade-off mechanism casts its signature by reducing or increasing the density of a group of circular phase islands that appear in each diffractive layer, which collectively help the image formation process by having a laterally aligned phase array, one diffractive layer following the next one. We also found out that a larger $L_1$ during the training process increased the density of these circular phase islands per optimized diffractive layer, which improved the general imaging quality of the diffractive network, but reduced its resilience to random distortions due to unknown diffusers.

We also demonstrated that an increased number (*n*) of training diffusers used in each epoch would help improve the network's generalization ability, especially for imaging through diffusers with larger $L_1$. Additionally, the overfitting of trained diffractive networks to known diffusers was also mitigated by increasing the total number (*N*) of training diffusers used through all the epochs. Although diffractive networks are composed of linear optical materials, their generalization capability is improved by increasing the number of diffractive layers, providing better image reconstruction performance through random diffusers. Finally, we also demonstrated a 'vaccination' method to increase the resilience of the diffractive network's image reconstruction performance to potential misalignments; we achieved this misalignment resilient imaging performance by introducing random lateral and axial shifts on successive diffractive layers during the training phase [28]. The overall analysis provided in this work presents a comprehensive guide for designing various computational imaging systems based on diffractive networks and might be useful for numerous fields, including, e.g., biomedical imaging, microscopy, atmospheric sciences, autonomous vehicles, and robotics.

## Results and discussions

### Design of diffractive neural networks for seeing through random diffusers

Various diffractive neural networks were designed to see through random, unknown diffusers to further explore the design space of this all-optical computational imaging platform. A randomly-selected amplitude-encoded object to be imaged was placed at 53λ in front of a random phase diffuser, where λ is the wavelength of the illumination source. The diffractive networks were composed of 4 successive trainable diffractive layers (Fig. 1), with an axial distance of 2.7λ in between. The output image plane was designed to be 9.3λ away from the last diffractive layer. To build the generalization ability to see through unknown random diffusers, we introduced multiple random diffusers in the training stage of each diffractive network. During the training phase, *n* uniquely different phase diffusers, each with a correlation length of $L_I$, were randomly generated to start an epoch. In each training iteration, a batch of *B* objects from the MNIST handwritten digit dataset was separately propagated to the diffuser plane. The optical field of each object was numerically duplicated into *n* different channels, one for each random phase diffuser. The resulting *B*×*n* optical fields were independently forward propagated through successive diffractive layers to reach the output plane, where their intensity profiles were collected and compared with the distortion-free input handwritten digits. A training cost function combining a structural loss term and energy efficiency penalty was used to train each diffractive network (see the Methods section). An epoch is marked as finished when all the 55k images in the MNIST dataset were exhausted. The training stopped after 100 epochs when the diffractive network has 'seen' *N*=100*n* different diffusers. After its training, the converged diffractive network can blindly image unknown new objects through both the known diffusers (used during the training phase) as well as new random diffusers that were never seen by the network. The Pearson Correlation Coefficient (PCC) [16] was used as our figure of merit for quantifying the quality of the all-optically reconstructed $D^2NN$ images of unknown objects hidden through unknown random diffusers.

**Generalization of diffractive neural networks to unknown random diffusers**

To explore the generalization ability of diffractive networks for imaging through random diffusers,

we first designed six different diffractive networks using training diffusers with correlation lengths $L_1$ of 5λ, 7.5λ, 10λ, 12.5λ, 15λ, and 17.5λ, and blindly tested them by hiding the objects behind random, new diffusers with a correlation length of $L_2$, where $L_1$ and $L_2$ are not necessarily equal to each other (Fig. 2(a)). Each one of these trained diffractive networks occupied a 'fading memory' that strongly overfitted to the last *n* diffusers ([$N-n+1 : N$] diffusers) used in training while also treating earlier training diffusers similar to randomly selected new diffusers[16]. To demonstrate and quantify this behavior, the average PCC values corresponding to the image reconstructions performed through the last *n* diffusers (blue crosses in Fig 2(b), $PCC_N$), the former *n* diffusers ([$N-2n+1 : N-n$] diffusers, red crosses in Fig 2(b), $PCC_{N-1}$), and 20 new randomly selected diffusers with $L_2=L_1$ (green circles, $PCC_{test}$) are compared in Fig. 2(b). An apparent increase in the reconstruction PCC values can be observed when $L_1$ is increased during the training. Furthermore, the performance gap between the average PCC values of known and new diffusers increased as $L_1$ was reduced. For example, for the diffractive network trained with $L_1$=17.5λ random diffusers, $PCC_N$ was calculated to be 0.866, while $PCC_{N-1}$=0.850 and $PCC_{test}$=0.847 when tested with $L_2=L_1$=17.5λ. On the other hand, the diffractive network trained with finer random diffusers ($L_1$=5λ) yielded $PCC_N$=0.656, $PCC_{N-1}$=0.608, and $PCC_{test}$=0.545 when tested with $L_2=L_1$=5λ. This increased performance gap reflects the challenge generated by finer phase diffusers with a smaller correlation length, which is also evident in the lens-based imaging results summarized in the last column of Fig. 2(c).

A diffractive network trained with $L_1$ diffusers can also be used to image objects hidden by new diffusers with a different correlation length $L_2$, i.e., $L_1 \neq L_2$. We tested each of our trained diffractive networks ($L_1$ = 5λ, 7.5λ, 10λ, 12.5λ, 15λ, and 17.5λ) with six different groups of new random phase diffusers, $L_2$=5λ, 7.5λ, 10λ, 12.5λ, 15λ, and 17.5λ, using 20 randomly selected diffusers in each group. The average $D^2NN$ image reconstruction PCC values over these test diffusers are reported in green circles in Fig. 2(b). Example images at the output FOV of the corresponding $D^2NN$ are also displayed in Fig. 2(c) to visualize the all-optical reconstruction fidelity.

All these trained diffractive networks, regardless of the value of $L_1$ used in training, provided a better image reconstruction when $L_2$ increases, evident from the PCC curves and the example images shown in Fig. 2. This improvement is in line with the fact that diffusers with larger $L_2$ values generate less distortion, creating a simpler reconstruction task for the diffractive network. In the same figure, we also showed the diffractive network reconstruction results when no diffuser is present. In fact, the best image reconstruction quality for each diffractive design was achieved without diffusers being present, illustrating that each diffractive design converged to a general-purpose imager, resilient to phase distortions due to random diffusers.

If we pay attention to each curve reported in Fig. 2(b), compared with $PCC_{test}$ values under $L_2=L_1$, a marginal improvement was observed when $L_2>L_1$, but a drastic drop appeared when $L_2<L_1$. The image reconstruction quality did not improve much when testing with $L_2>L_1$, while the speckle noise increased, and the reconstructions became worse when $L_2<L_1$. This behavior indicates that the trained diffractive networks suffer a trade-off between image reconstruction fidelity and distortion reduction capability. To shed more light on the mechanism of this trade-off, we examined the converged diffractive networks trained with different $L_1$ values (see Fig. 3). A characteristic phase pattern was observed for all the designed diffractive layers, even though they were trained with diffusers of varying $L_1$ values. These spatial patterns consisted of relatively smooth, circular phase islands surrounded by rapidly changing diffractive neurons, which collectively contributed to the reconstruction of the images distorted by random diffusers. The specific functions of these circular phase islands are better illustrated in Fig. 3, where we reconstructed an unknown object without a diffuser, through a known diffuser and through a new random diffuser ($L_2=L_1$) using different diffractive networks trained with $L_1$=5λ, 10λ, and 15λ; in different rows for each case shown in Fig. 3, we applied different levels of pruning to isolate different spatial features to allow them to work separately. As shown in the first rows of Figs. 3(a)-(c), all the trained diffractive networks can image unknown objects through both known and new diffusers and also successfully reconstruct an image when the diffuser is absent. In the second rows of Figs. 3(a)-(c), we show the results when we pruned all the diffractive neurons outside of the circular phase islands in each diffractive layer, i.e., retaining the modulation of the circular phase islands and setting other regions of the diffractive layer to have zero relative phase modulation. Using only the circular phase islands, all three diffractive networks can still image the object without any diffuser present. This indicates that the main function of these circular phase islands that are laterally self-aligned (matching their relative positions across different layers) is image formation between the input and output FOVs of the diffractive network. On the other hand, using only these circular phase islands per diffractive later, the image reconstruction performance through random diffusers got worse, as illustrated in Fig. 3. Reducing the level of pruning to include the diffractive neurons outside of the phase islands improved the diffractive network's resilience to distortions generated by random diffusers, as illustrated in the third rows of Figs. 3(a)-(c). By keeping all the diffractive features within a circular aperture of 80λ on each layer, the image reconstruction fidelity improved to a level close to the original diffractive design, where all the neurons are accessible. Based on this analysis detailed above, we can conclude that the self-aligned circular phase islands and the rapidly changing diffractive features surrounding them have different functions in a diffractive imager: the circular phase islands aim to map the object/input FOV of the diffractive network to the output FOV, whereas the rapidly changing diffractive phase features around these phase islands aim to get rid of the distortions caused by random diffusers, by coupling the electromagnetic waves that mainly carry the distortion information outside of the receptive field of the successive diffractive layer(s).

Another interesting observation that one can make is that the $L_1$ values used in training guided the diffractive network designs to distribute and self-align their circular phase islands differently. To shed more light, we counted the number of these self-aligned phase islands on each diffractive layer (see Methods section for details) for different $L_1$ values. For each diffractive network design, the mean and the standard deviation of the number of phase islands among its four diffractive layers were calculated (plotted in Fig. 4(a)). Based on this analysis, we observe that the number of phase islands increases with larger $L_1$, which implies that a diffractive network needs less of the rapidly changing diffractive phase features for imaging weakly distorted objects under a large correlation length. In this case, the diffractive network optimization prefers to use its diffractive neuron budget more for building self-aligned phase islands that are densely packed per diffractive layer to improve its image quality at the output. In the opposite case of using smaller $L_1$ values during the training, the diffractive networks are optimized to mitigate and suppress the wave distortion caused by random phase diffusers, and therefore the network prefers to have more of the rapidly changing diffractive phase features at the cost of reducing the density of the self-aligned phase islands per layer.

To better quantify this trade-off mechanism, we calculated the diffractive mask vacancy ratio (see Fig. 4 and the Methods section), which describes the percentage of the diffractive pixels/neurons that do not belong to the self-aligned phase islands on each diffractive layer. Figure 4 illustrates that a larger $L_I$ during the training phase leads to diffractive layer designs with smaller but denser phase islands aiming to improve the image quality at the output; on the other hand, a smaller $L_I$ during the training leads to an increase in the mask vacancy since the network prefers to have more of the rapidly changing diffractive phase features per layer to couple out the distorted waves and mitigate the presence of the stronger random phase diffusers hiding the objects.

**Improved training strategies with better generalization**

Each of the diffractive networks reported in the previous sections was trained with random diffusers following the same statistical distribution, i.e., the correlation lengths $L_I$ for all the training diffusers were kept the same. Diversifying the training diffusers to cover a set of different $L_I$ values can improve the generalization ability of the resulting diffractive network. To explore this, we implemented two methods: the first was to use random training diffusers with several distinct $L_I$ values. As an example, we trained diffractive networks with $L_I \in \{7.5\lambda, 10\lambda, 12.5\lambda\}$ and $L_I \in \{10\lambda, 12.5\lambda, 15\lambda\}$. In the second method, we trained networks with random diffusers whose correlation lengths were randomly selected from a uniform distribution, i.e., $L_I \in U[7.5\lambda, 12.5\lambda]$, $L_I \in U[10\lambda, 15\lambda]$, and $L_I \in U[5\lambda, 17.5\lambda]$; see the Methods section for details. After being trained for 100 epochs, we analyzed the image reconstruction performance of each one of these diffractive networks for seeing through known and new random diffusers, the results of which are shown in Fig. 5. Compared to the diffractive networks trained with a single $L_1$ value, for example, $L_1$=10λ, using diffusers with both finer and coarser grain sizes during the training phase provided a better balance between the output image quality and distortion reduction. The "vanilla" diffractive network trained with $L_1$=10λ yielded a $PCC_{test}$=0.709 for $L_2$=$L_1$, while the networks trained with $L_1 \in$ {7.5λ, 10λ, 12.5λ} ($L_1 \in$U[7.5λ, 12.5λ]) achieved $PCC_{test}$ values of 0.733 (0.743) when tested with $L_2$=10λ random diffusers, demonstrating better image reconstruction performance. More importantly, this new training strategy also improved the reconstruction fidelity for seeing through random diffusers that are outside the range of $L_1$. The original vanilla diffractive network trained with $L_1$=10λ yielded $PCC_{test}$=0.503 when tested with $L_2$=5λ random diffusers and $PCC_{test}$=0.757 for $L_2$=17.5λ. These PCC values raised to 0.515/0.520 ($L_2$=5λ) and 0.783/0.786 ($L_2$=17.5λ) using diffractive networks trained with $L_1 \in$ {7.5λ, 10λ, 12.5λ}/ $L_1 \in$U[7.5λ, 12.5λ], respectively. As another example, using the diffractive network trained with $L_1$=10λ as a baseline, broadening $L_1$ to cover up to 15λ diffusers during the training phase raised the $PCC_{test}$ ($L_2$=17.5λ) from 0.757 to 0.820 or 0.823 using $L_1 \in$ {10λ, 12.5λ, 15λ} or $L_1 \in$U[10λ, 15λ], respectively. The $PCC_{test}$ under $L_2$=5λ random diffusers, however, dropped from 0.503 to 0.493 and 0.489, using $L_1 \in$ {10λ, 12.5λ, 15λ} or $L_1 \in$U[10λ, 15λ], respectively.

**An increase in the number of random training diffusers helps generalization**

Another degree of freedom that can be used to optimize a network's generalization performance is the number of diffusers $n$ used in each training epoch. To explore the impact of $n$, we first trained three diffractive networks using $n$=10, 20, and 40 diffusers ($L_1$=10λ) and blindly tested them using randomly selected diffusers with various $L_2$ values (first three columns in Fig. 6(a)). The total number of random diffusers used in the training stage ($N$) was kept unchanged, i.e., $N$ = 2000. Therefore, the total training epochs were 200, 100, and 50 for these three diffractive networks trained using $n$=10, 20, and 40 diffusers, respectively. This varying number of training epochs led to a bias since the network trained with a smaller $n$ achieved a better imaging performance through $L_2$>10λ diffusers; this is expected since the networks with a smaller $n$ went through more optimization steps through a larger number of epochs. Next, we used $n$=40 diffusers in each epoch and trained the corresponding diffractive network for 100 epochs (fourth column in Fig. 6(a)), i.e., $N$=4000 randomly selected training diffusers were used in total to form a fair comparison. A further performance improvement for imaging through new diffusers with various $L_2$ values was observed against the $n$=40, $N$=2000 diffractive network. The positive impact of $n$ can also be seen by comparing $n$=40, $N$=4000 diffractive network results with those of $n$=20, $N$=2000 diffractive network (Fig. 6).

Same as what we observed in the previous subsections, the diffractive networks trained in these examples still 'remembered' the last [$N$-$n$+1: $N$] diffusers, achieving better image reconstruction through those known diffusers of the last epoch compared to new, unknown diffusers (see the blue crosses in Fig. 6(a)). To shed more light on this, we quantified this memory effect by defining a PCC overfitting rate as:

$$PCC\ Overfitting\ Rate = \frac{PCC_N - PCC_{test}}{PCC_N},$$

where $PCC_{test}$ is calculated for the new random diffusers with $L_2$= $L_1$=10λ (Fig. 6(c)). As illustrated in Fig. 6(c), the PCC overfitting rate decreased with a larger $n$ when $N$ was kept the same, and a larger $N$ further decreased the PCC overfitting rate for the same $n$, illustrating the improved generalization capability of the diffractive network.

**Training initialization**

The diffractive networks reported in the previous subsections were trained from scratch, where each diffractive layer had a uniform, zero-phase modulation at the beginning/initialization. The initialization strategy is in general important for successful training of digital neural networks[29], [30], and similarly for diffractive networks[18], [31], [32]. To evaluate the impact of diffractive layer initialization, we first trained a diffractive network from scratch to image handwritten digits without any diffusers. The diffractive layers obtained after 50 epochs were then used to initialize the optimization of a second diffractive network that was trained for another 100 epochs to image through random phase diffusers (see Figs. 7(a) and 7(b) for the resulting diffractive layers). The diffractive layers after the transfer learning step look very similar to the resulting diffractive layers with zero phase initialization, showing similar characteristic patterns with the circular phase islands surrounded by the high spatial frequency diffractive neurons (see Fig. 7(b)). Despite this similarity, the transfer-learned phase islands appear sparser compared to the diffractive layers from zero-phase initialization. However, this sparsity of the circular phase islands in each diffractive layer did not degrade the imaging performance. In contrast, the diffractive network that is transfer-learned from a diffuser-free design achieved a slightly better imaging performance, as illustrated in Fig. 7(c). This means that the prior knowledge of a diffractive imaging system that is transfer-learned as an initial condition guided the diffractive network optimization to arrive at a more optimal solution to image objects hidden through unknown random diffusers, exemplifying the importance of the diffractive layer initialization during the learning process.

**Impact of the network depth: Imaging performance through unknown random diffusers improves for deeper diffractive networks**

Next, we compared the generalization ability of diffractive networks with different numbers of diffractive layers (see Fig. 8). All these diffractive networks were trained with $L_1$=10λ diffusers and tested with different $L_2$ values, and the corresponding image reconstruction PCC values of each diffractive network are reported in Fig. 8(a). These results confirm the depth advantage of diffractive networks: single-layer and two-layer networks have relatively low PCC values in their image reconstruction, and the imaging performance improves as we increase the number of diffractive layers. For example, for the most challenging case of $L_2$=5λ, the reconstructed images of one-, two-, or three-layer diffractive networks were hard to distinguish the objects due to strong speckle patterns. However, using four or five diffractive layers, handwritten digits were recognizable, although these diffractive networks were only trained with $L_1$=10λ diffusers. On the other hand, for the single-layer and two-layer diffractive networks, even the easiest task of imaging objects through $L_2$=17.5λ random diffusers turned out to be challenging, as illustrated in Fig. 8. These results indicate the depth advantage of diffractive networks that favor distributing the available diffractive neurons into deeper architectures, one layer following another[33].

**Mitigating misalignments through diffractive network vaccination**

Experimental demonstrations of deeper diffractive networks can be challenging due to fabrication inaccuracies and mechanical misalignments. To mitigate potential misalignments and their negative impact on diffractive inference, we adopted a training strategy termed "vaccination"[28], which randomly shifts the diffractive layers (on purpose) during the training process to increase the resilience of the diffractive network against random physical misalignments. To implement vaccination for imaging through random diffusers, we introduced a uniformly distributed random 3D displacement vector ($\boldsymbol{D} = (D_x, D_y, D_z)$) for each diffractive layer, and designed four different vaccinated diffractive networks with $\boldsymbol{L_1}$=10λ using varying levels of layer displacements; see Fig. 9(a). The maximum amount of mechanical shift allowed along the corresponding axes was selected as $\boldsymbol{\Delta}_{x,tr} = \boldsymbol{\Delta}_{y,tr} = \boldsymbol{\Delta}_{z,tr} = $ 0λ, 0.5λ, 1.0λ, and 2.0λ, respectively, i.e., $D_x \sim \mathbf{U}(-\Delta_{x,tr}, \Delta_{x,tr})$, $D_y \sim \mathbf{U}(-\Delta_{y,tr}, \Delta_{y,tr})$ and $D_z \sim \mathbf{U}(-\Delta_{z,tr}, \Delta_{z,tr})$; see the Methods section for details. After the training process, the vaccinated models were blindly tested with multiple levels of random displacements and the image reconstruction fidelity of these vaccinated diffractive networks to image objects through $\boldsymbol{L_2}$=10λ diffusers is reported in Fig. 9(a). For the diffractive network trained without any random physical shifts (i.e., non-vaccinated), a quick drop in the image reconstruction quality appears when random shifts are introduced in the positions of the diffractive layers. For example, for the non-vaccinated diffractive network, the all-optically reconstructed images of handwritten digit '3' are hard to recognize due to speckle patterns for $\Delta_{test}$ larger than 1λ (see the first column in Fig. 9(b)). As for the diffractive network trained with $\boldsymbol{\Delta}_{x,tr} = \boldsymbol{\Delta}_{y,tr} = \boldsymbol{\Delta}_{z,tr} = 1\lambda$, the image reconstruction was more robust to misalignments (third column in Fig. 9(b)), maintaining acceptable imaging performance even if the testing shifts were much larger than the training. A similar misalignment resilience was also observed for the diffractive network trained with $\boldsymbol{\Delta}_{x,tr} = \boldsymbol{\Delta}_{y,tr} = \boldsymbol{\Delta}_{z,tr} = 2\lambda$ (see Fig. 9).

# Methods

### Random diffuser design

We used a phase-only mask to model a random phase diffuser, whose transmittance $t_D(x, y)$ is defined by the refractive index difference ($\Delta n$) between air and the diffuser material, and a random height-map $D(x, y)$ at the diffuser plane, i.e.,

$$t_D(x, y) = exp\left(j \frac{2\pi \Delta n}{\lambda} D(x, y)\right) \quad (1)$$

where $j = \sqrt{-1}$ and $\lambda$ is the illumination wavelength. The random height-map $D(x, y)$ is defined as

$$D(x, y) = W(x, y) * K(\sigma) \quad (2)$$

where $W(x, y)$ follows a normal distribution with a mean µ and a standard deviation $\sigma_0$, i.e.

$$W(x, y) \sim \mathcal{N}(\mu, \sigma_0^2) \quad (3)$$

$K(\sigma)$ is a zero-mean Gaussian smoothing kernel with a standard deviation of $\sigma$, and '$*$' denotes the 2D convolution operation. We can calculate the mean correlation length ($L$) of each diffuser using a phase-autocorrelation function $R_d(x, y)$ defined as:

$$R_d(x, y) = \exp(-\pi(x^2 + y^2)/L^2) \quad (4)$$

In this work, for µ = 25$\lambda$, $\sigma_0 = 8\lambda$ and $\sigma = 4\lambda$, we verified the average correlation length as $L \sim 10\lambda$ based on 2000 randomly generated diffusers using the phase-autocorrelation function. We accordingly modified the σ values to generate the corresponding diffusers for the other correlation lengths used in this work.

We used three strategies to train our diffractive optical networks: (*i*) using a fixed *L₁* for all the random diffusers, (*ii*) using *L₁* selected from a set of discrete values, and (*iii*) using *L₁* selected uniformly from a continuous range of values. For the training strategy using three discrete $L_1$ values, each one of the $n$ diffusers was randomly assigned with a correlation length. For those designs where $L_1$ covers a continuous range, we first split the range into $n$ uniform intervals and selected a correlation length from each interval uniformly to generate the corresponding diffuser. The sequence of the generated random diffusers was shuffled, and therefore the diffusers' correlation lengths were not monotonically increasing or decreasing during the optimization iterations.

**Forward propagation model**

Free space light propagation between the diffractive layers was calculated using the Rayleigh-Sommerfeld equation[17], which can be expressed as:

$$w(x, y, z) = \frac{z}{r^2}\left(\frac{1}{2\pi r^2} + \frac{1}{j\lambda}\right) exp\left(\frac{j2\pi r}{\lambda}\right) \quad (5)$$

where $r = \sqrt{x^2 + y^2 + z^2}$.

A random phase diffuser located at $z_0$ introduces a phase distortion $t_D(x, y)$. Considering a plane wave that was incident at an amplitude-modulated image $h(x, y, z = 0)$ positioned at $z = 0$, we formulated the distorted image right after the diffuser as:

$$u_0(x, y, z_0) = t_D(x, y) \cdot [h(x, y, 0) * w(x, y, z_0)] \quad (6)$$

This distorted field was used as the input field of subsequent diffractive layers. We modeled the diffractive layers as thin phase elements, and the transmittance of layer $m$ located at $z = z_m$ can be formulated as:

$$t_m = exp(j\phi(x, y, z_m)) \quad (7)$$

The optical field $u_m(x, y, z_m)$ right after the $m^{th}$ the diffractive layer at $z = z_m$ can be

written as:

$$u_m(x, y, z_m) = t_m(x, y, z_m) \cdot [u_{m-1}(x, y, z_{m-1}) * w(x, y, \Delta z_m)] \quad (8)$$

where $\Delta z_m = z_m - z_{m-1}$ is the axial distance between two successive diffractive layers, which was selected as $2.7\lambda$ in this paper. After being modulated by all the $M$ diffractive layers, the optical field was collected at an output plane which was $\Delta z_d = 9.3\lambda$ away from the last diffractive layer. Then, the intensity of the optical field was used as the output of the network:

$$o(x, y) = |u_M * w(x, y, \Delta z_d)|^2 \quad (9)$$

**Numerical implementation**

The input field of the diffractive neural networks in this paper was assumed to be a coherent illumination with a wavelength of $\lambda \approx 0.75 \ mm$. Each diffractive layer contained $240 \times 240$ diffractive neurons with a pixel size of 0.3 mm and only modulated the phase of the incident light field. During the training process, samples from the MNIST training dataset were first bilinearly interpolated from $28 \times 28$ pixels to $160 \times 160$, and padded with zeros to $240 \times 240$ pixels. In each training batch, $B = 10$ different MNIST images were sampled randomly. Besides, each input object $h_b(x, y)$ in a batch was numerically duplicated $n$ times and separately perturbed by a set of $n$ randomly selected diffusers. Therefore, $B \times n$ different optical fields were obtained, and these distorted fields were individually forward propagated through the diffractive network. At the output plane, which was $160 \times 160$ pixels large, we got $B \times n$ different intensity patterns: $o_{11}, \ldots, o_{Bn}$, which were used for the loss function calculation:

$$Loss = \frac{\sum_{b,i=1,1}^{b=B, i=n}[-P(o_{bi}, h_b) + E(o_{bi}, h_b)]}{B \times n} \quad (10)$$

where $P(o_{bi}, h_b)$ is the PCC between the output intensity image and its ground truth image $h_b$.

$E(o_{bi}, h_b)$ denotes an energy efficiency-related regularization term, defined as:

$$E(o_{bi}, h_b) = \frac{\sum_{x,y}\left(\alpha(1 - \widehat{h_b}) \cdot o_{bi} - \beta \widehat{h_b} \cdot o_{bi}\right)}{\sum_{x,y} \widehat{h_b}} \quad (11)$$

where $\widehat{h_b}$ is a binary mask indicating the transmittance area on the input object, defined as:

$$\widehat{h_b}(x, y) = \begin{cases} 1, & for\ (x, y)\ if\ h_b(x, y) > 0 \\ 0, & otherwise \end{cases} \quad (12)$$

where α and β are hyper-parameters set to be 1 and 0.5, respectively.

The calculated loss value was then backpropagated to update the pixel phase modulation values using the Adam optimizer[34] with a decaying learning rate of $Lr = 0.99^{epoch} \times 10^{-3}$, where $epoch$ refers to the current epoch number.

Our models were trained using Python (v3.6.13) and TensorFlow (v1.15.0, Google Inc.) with a GeForce GTX 1080 Ti graphical processing unit (GPU, Nvidia Inc.), an Intel® Core™ i7-7700K central processing unit (CPU, Intel Inc.) and 64 GB of RAM, running the Windows 10 operating system (Microsoft). Training a typical diffractive neural network model takes ~36 h to complete with 100 epochs and *n*=20 diffusers per epoch.

**Phase island extraction and mask vacancy ratio calculation**

We created a MATLAB program to extract the self-aligned circular phase islands in each layer. We first applied a *cosine* function to the phase profile. Then we binarized the image and performed the morphological close operation to get the outlines of the phase islands. The mis-detected regions were manually removed. For the phase island number calculation, we applied morphological erosion to the previously calculated binary result to separate the phase islands. The number and size of the phase islands were calculated using conventional connected component analysis. The mask vacancy ratio ($V_{L_1}$) over four layers of a diffractive neural network designed with $L_1$ diffusers was calculated as:

$$V_{L_1} = \sum_{j=1}^{4}\left(1 - \frac{\sum_{i \in P_{L_1,j}} s_i}{240 \times 240}\right) / \sum_{j=1}^{4}\left(1 - \frac{\sum_{i \in P_{10\lambda,j}} s_i}{240 \times 240}\right) \quad (13).$$

where $P_{L_1,j}$ denotes all the phase islands detected on the $j^{th}$ layer of a diffractive neural network trained with $L_1$ diffusers, each of which has $s_i$ pixels.

**Image contrast enhancement**

We digitally enhanced the contrast of all the images using a built-in MATLAB function (*imadjust*) for visualization. By default, the function saturated the bottom 1% and the top 1% of all pixel values. Besides, all the quantitative analyses were based on raw image data.

**Lens-based imaging system simulation**

We simulated a Fresnel lens-based imaging system serving as a comparison to evaluate how a random diffuser affects the output image; see, e.g., Fig. 1(b). The designed lens had a focal length (*f*) of 145.6λ and a pupil diameter of 104λ. The transmission coefficient of the lens $t_L$ was formulated as:

$$t_L(\Delta x, \Delta y) = A(\Delta x, \Delta y) exp\left(-j\frac{\pi}{\lambda f}(\Delta x^2 + \Delta y^2)\right) \quad (14)$$

where $\Delta x$ and $\Delta y$ denote the distance from the center of the lens in lateral coordinates. $A(\Delta x, \Delta y)$ is the amplitude function, i.e.,

$$A(\Delta x, \Delta y) = \begin{cases} 1, & \sqrt{\Delta x^2 + \Delta y^2} < 52\lambda \\ 0, & \text{otherwise} \end{cases} \quad (15)$$

The lens was $2f$ (291.2λ) away from the input object. Using the angular spectrum method,

the light from the input object was propagated to the random diffuser plane ($z_0 = 53\lambda$). Then the light field was distorted by the random phase diffuser and propagated to the lens plane. After passing through the lens, and applying the angular spectrum method again, the resulting complex field was propagated to the image plane ($2f$ behind the lens). The intensity pattern at the image plane was regarded as the output image, created by an aberration-free lens through a random phase diffuser.

**Vaccination**

During the training process of a vaccinated diffractive network, a uniformly distributed random 3D displacement $\boldsymbol{D} = (D_x, D_y, D_z)$ was added to each diffractive layer, i.e.,

$$D_x \sim \mathbf{U}(-\Delta_{x,tr}, \Delta_{x,tr})$$
$$D_y \sim \mathbf{U}(-\Delta_{y,tr}, \Delta_{y,tr}) \qquad (16)$$
$$D_z \sim \mathbf{U}(-\Delta_{z,tr}, \Delta_{z,tr})$$

where $D_x$ denotes the left-right displacement of a layer, $D_y$ denotes the up-down displacement of a layer, $D_z$ denotes the displacement of the layer-to-layer distance (i.e., $\Delta z_m$), and $\Delta_*$ represents the maximum amount of physical shift allowed along the corresponding axis, limiting the upper and lower bound of the uniform distribution. In the training process, the vaccination amount was selected to be $\Delta_{x,tr} = \Delta_{y,tr} = \Delta_{z,tr} = 0$, $\Delta_{x,tr} = \Delta_{y,tr} = \Delta_{z,tr} = \frac{1}{2}\lambda$, $\Delta_{x,tr} = \Delta_{y,tr} = \Delta_{z,tr} = \lambda$, and $\Delta_{x,tr} = \Delta_{y,tr} = \Delta_{z,tr} = 2\lambda$, respectively, representing different levels of estimated uncertainty during the experimental implementations. Note that $D_x$, $D_y$, and $D_z$ of each diffractive layer were independently sampled from the given uniform random distributions. Except for the different amounts of random misalignments, all the other parameters were kept the same throughout the training process.

After the training process was completed, the vaccinated models were blindly tested with multiple levels of displacements over the MNIST test images. During each test, a fixed displacement value $\Delta_{test}$ was applied to the system so that the lateral displacements $(D_x, D_y)$ of the four consecutive diffractive layers were set as $(-\Delta_{test}, -\Delta_{test})$, $(-\Delta_{test}, \Delta_{test})$, $(\Delta_{test}, \Delta_{test})$, and $(\Delta_{test}, -\Delta_{test})$ respectively to simulate the *worst-case* scenarios. The displacement of the layer-to-layer distance $D_z$ was set to $\Delta_{test}$. In other words, the four diffractive layers were laterally shifted toward the four corners, and the layer-to-layer distance between any two consecutive layers was increased by $\Delta_{test}$. Each diffractive network model was tested with $\Delta_{test} = 0$, $\Delta_{test} = \frac{1}{2}\lambda$, $\Delta_{test} = \lambda$, $\Delta_{test} = \frac{3}{2}\lambda$, and $\Delta_{test} = 2\lambda$, respectively. For each model tested with each level of displacement, an average PCC value between the ground truth and the output images was calculated to evaluate the model's performance when misalignments were present.

## Acknowledgments

The Ozcan Research Lab at UCLA acknowledges the support of ONR (Office of Naval Research).

**Figures and Figure Captions**

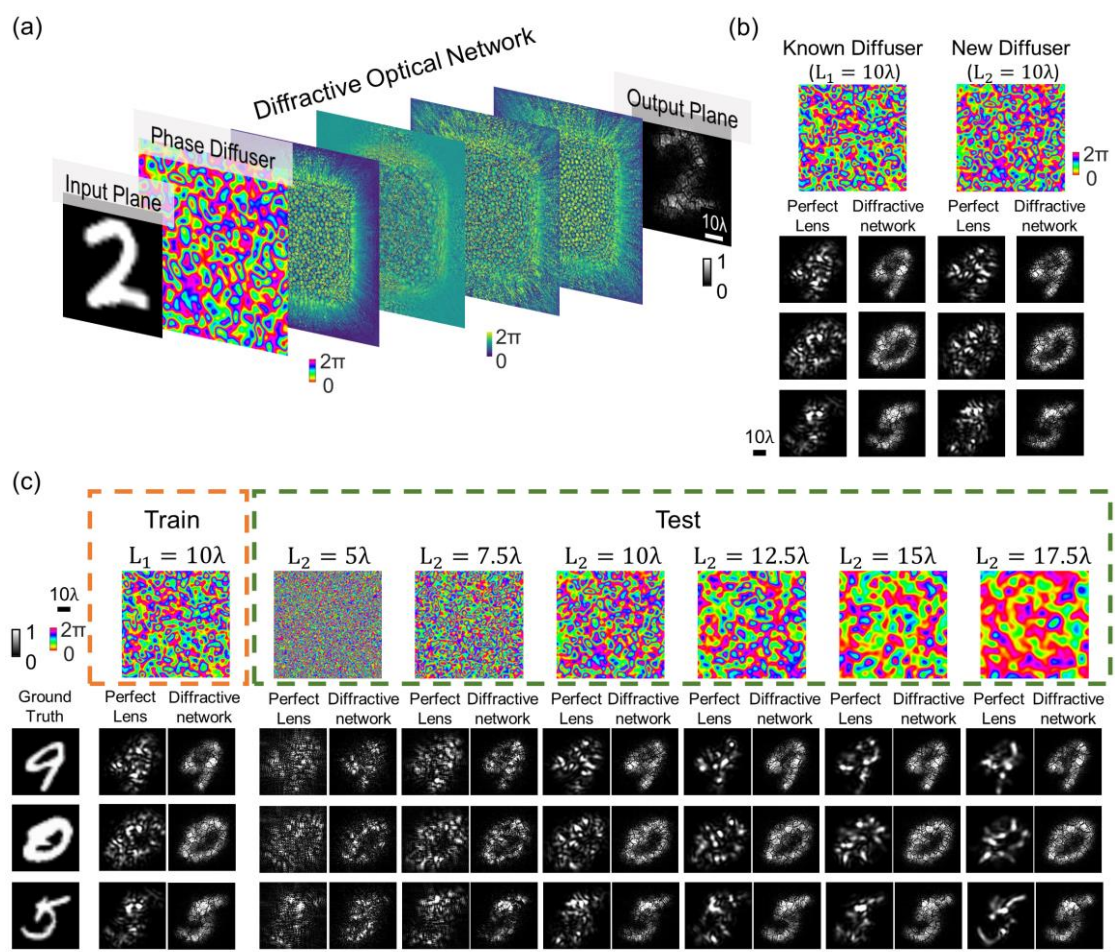

**Figure 1. All-optical imaging through diffusers using diffractive networks.** (a) Schematic of a four-layered diffractive network trained to all-optically reconstruct the input field-of-view seen through an unknown random diffuser. (b) Sample images, seen through known and new diffusers ($L_1=L_2=10\lambda$) using a perfect lens and a trained diffractive optical network. (c) Imaging unknown objects through unknown, new diffusers with correlation lengths of $L_2$=5λ, 7.5λ, 10λ, 12.5λ, 15λ, and 17.5λ, using a diffractive optical network trained with $L_1$=10λ. The contrast of the images was enhanced for visualization (see the Methods section for details).

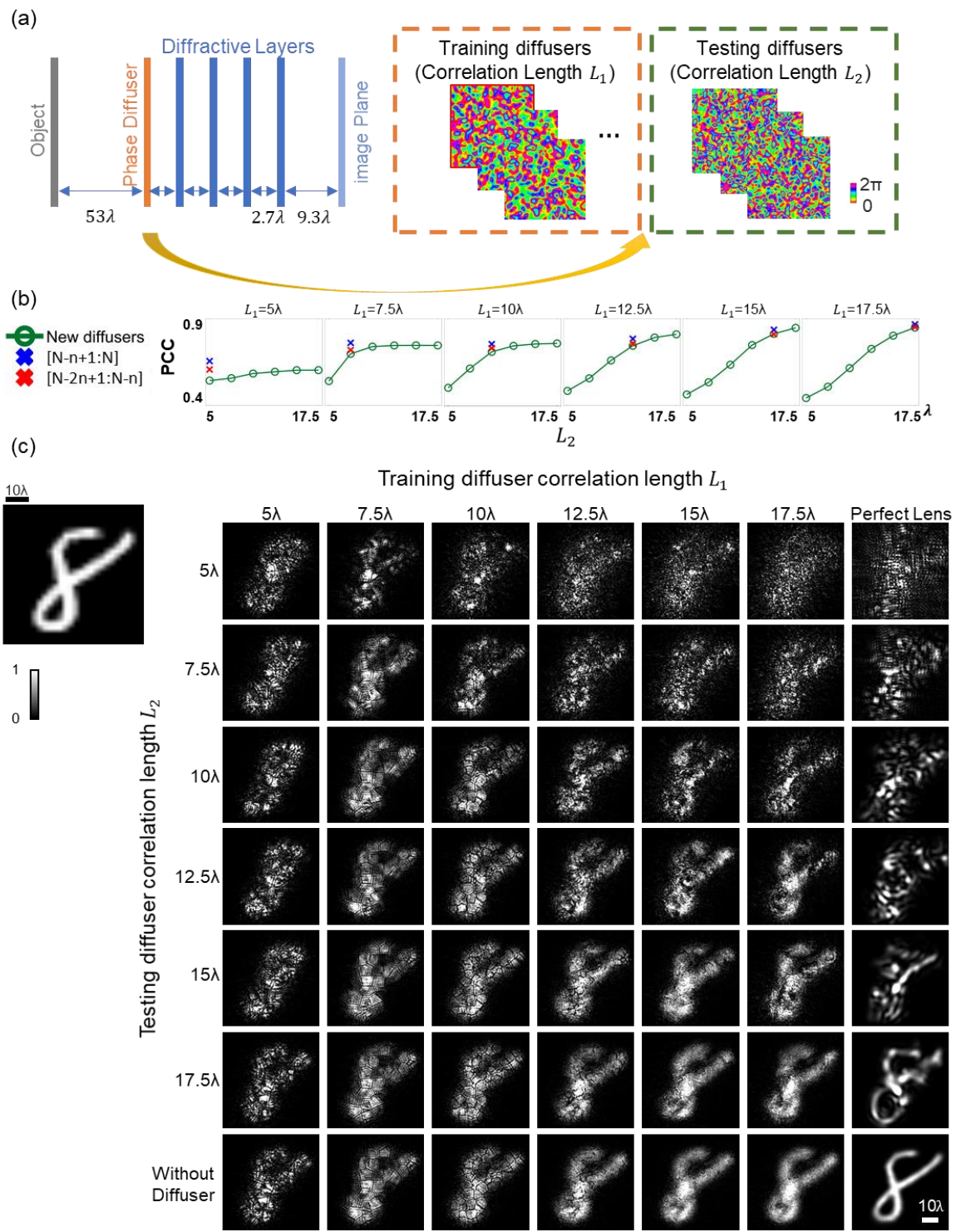

**Figure 2. Imaging through random diffusers with different correlation lengths.** (a)Training and design schematic of a 4-layered diffractive optical network seeing through new, random phase diffusers. (b) PCC values of diffractive neural networks trained with $L_1$=5λ, 7.5λ, 10λ, 12.5λ, 15λ,

and 17.5λ random diffusers and tested with new, random phase diffusers with correlation lengths $L_2$=5λ, 7.5λ, 10λ, 12.5λ, 15λ, and 17.5λ. (c) Visualization of the results in (b). The last column: imaging through the same random diffusers using a perfect lens. Last row: imaging without a diffuser. The contrast of the images was enhanced for visualization (see the Methods section for details).

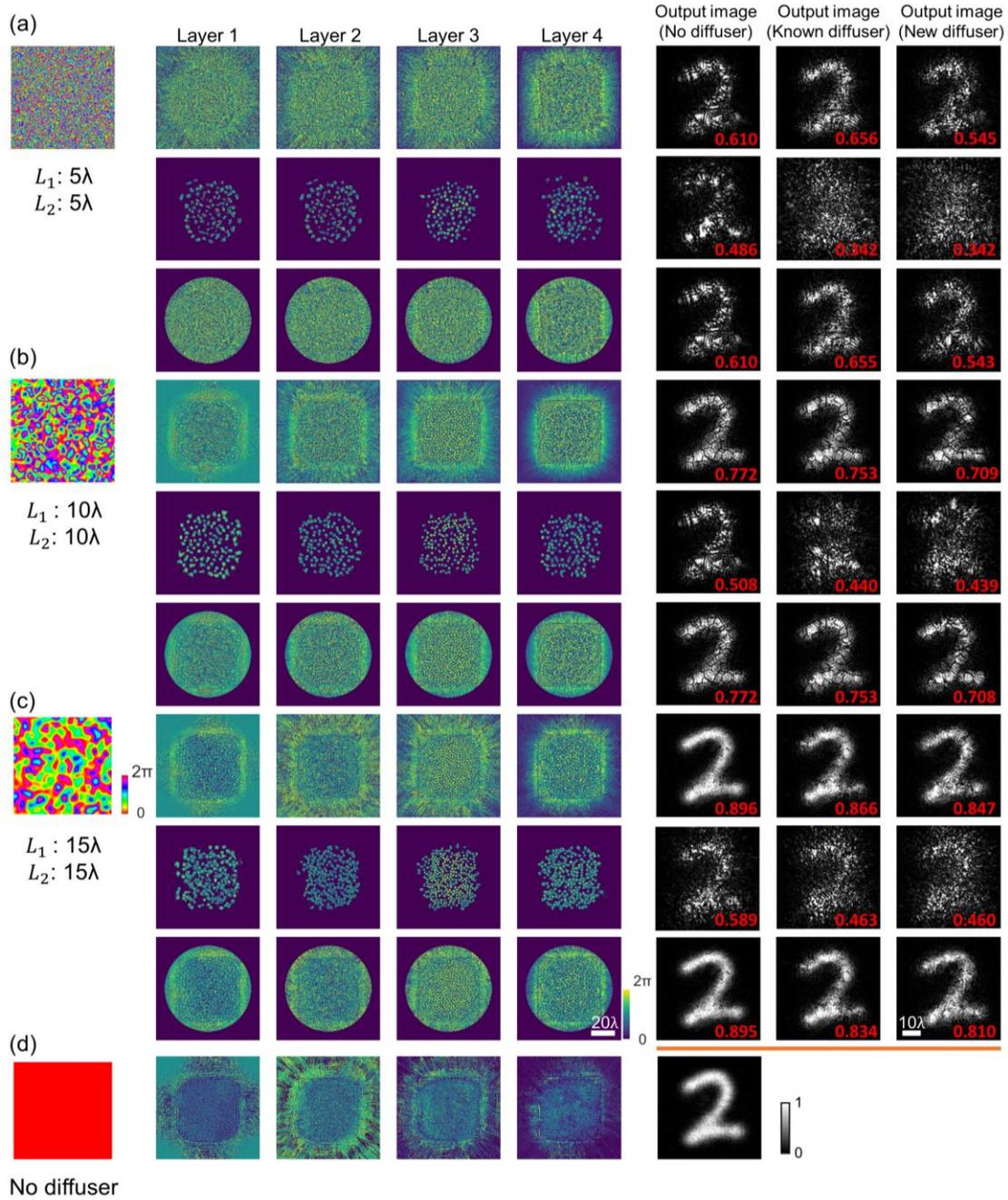

**Figure 3. Comparison of diffractive network output images under different levels of pruning.** The diffractive layers trained with correlation lengths of $L_I$=5λ (a), 10λ(b), and 15λ(c) were pruned by different levels. In each panel: top row: the original layers of a trained diffractive network and its reconstruction results under different conditions; second row: the diffractive layers where all the neurons outside of the phase islands are pruned; third row: diffractive layers where all the neurons

within a circular aperture of $80\lambda$ were kept. (d) The diffractive network trained and tested without any diffusers. The contrast of the images was enhanced for visualization (see the Methods section for details).

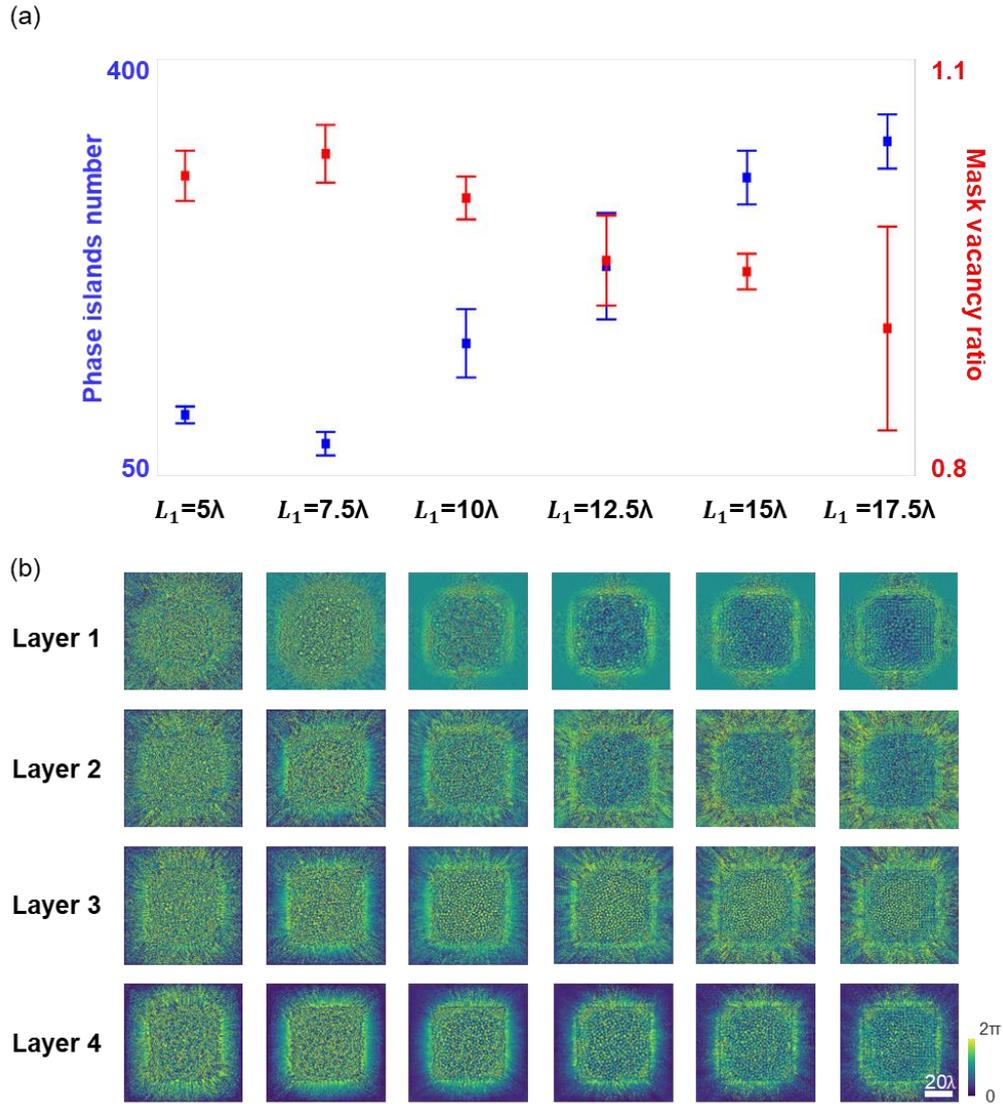

**Figure 4. Quantification of the circular phase islands within the diffractive layers.** (a) The number of the phase islands and the mask vacancy ratio of the diffractive layers trained with different random phase diffusers ($L_1 = 5\lambda$, $7.5\lambda$, $10\lambda$, $12.5\lambda$, $15\lambda$ and $17.5$). (b) Corresponding diffractive layer phase profiles of the converged diffractive networks.

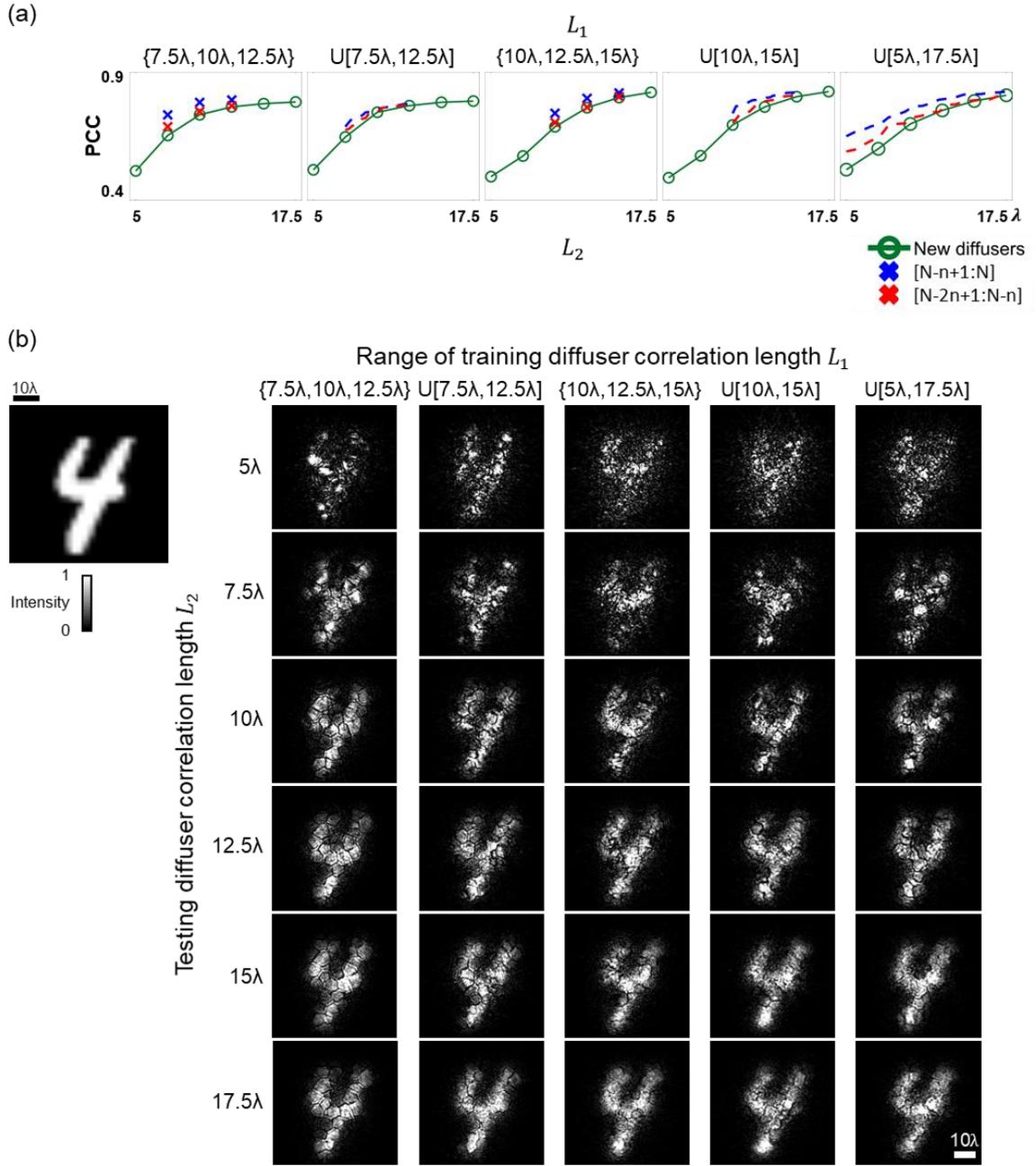

**Figure 5. Improved training strategies with better generalization.** (a) PCC values of various diffractive neural networks trained and tested with different $L_1$, $L_2$ combinations. (b) Example images seen through new random diffusers using the diffractive networks in (a). The contrast of the images was enhanced for visualization (see the Methods section for details).

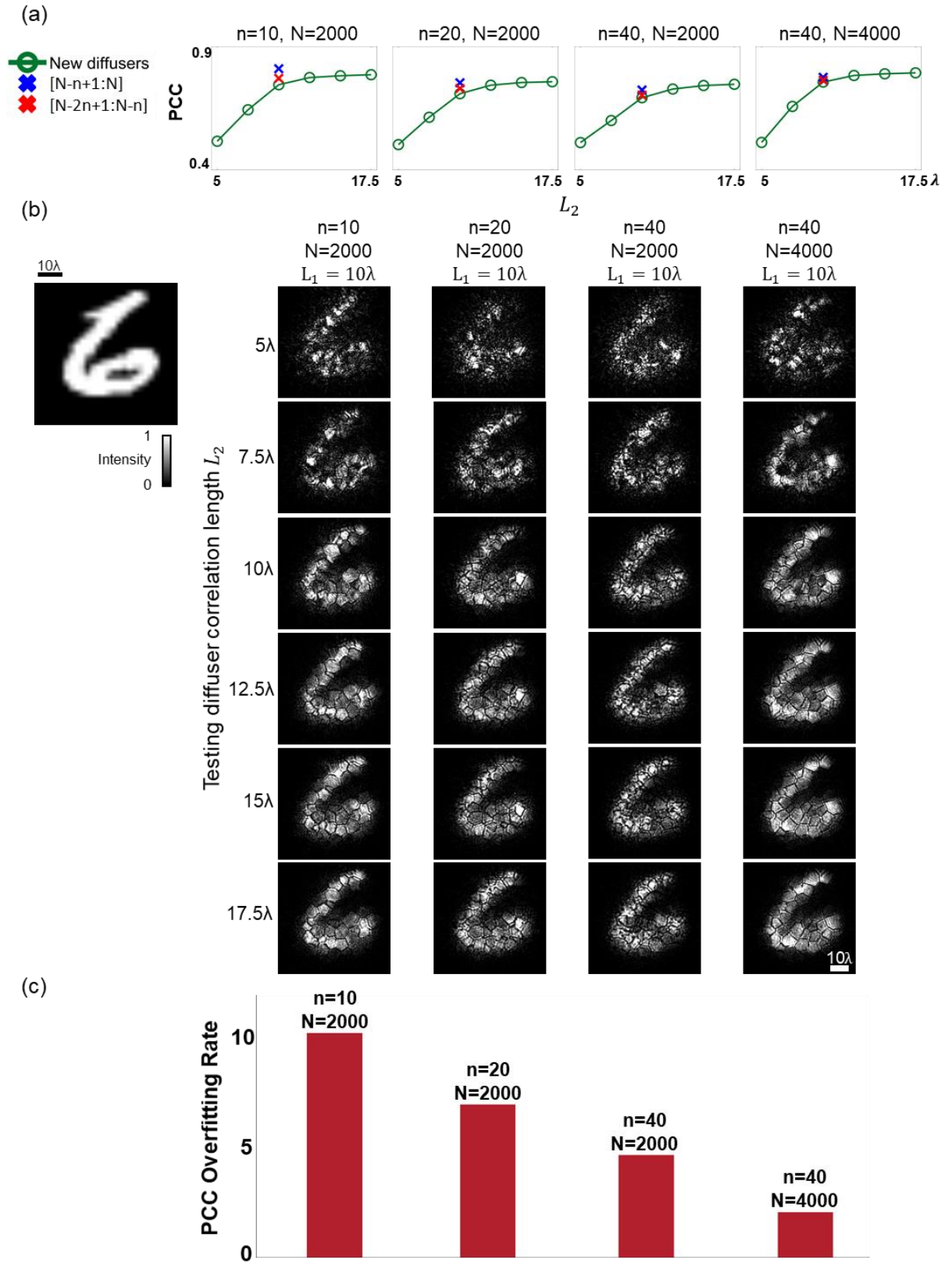

**Figure 6. An increase in the number of random training diffusers helps generalization.** (a) PCC

values of the diffractive neural networks trained with different combinations of *n* and *N*. (b) Example images seen through new random diffusers using the diffractive networks in (a). (c) PCC overfitting rate for the four diffractive networks trained with different combinations of *n* and *N*. The contrast of the images was enhanced for visualization (see the Methods section for details).

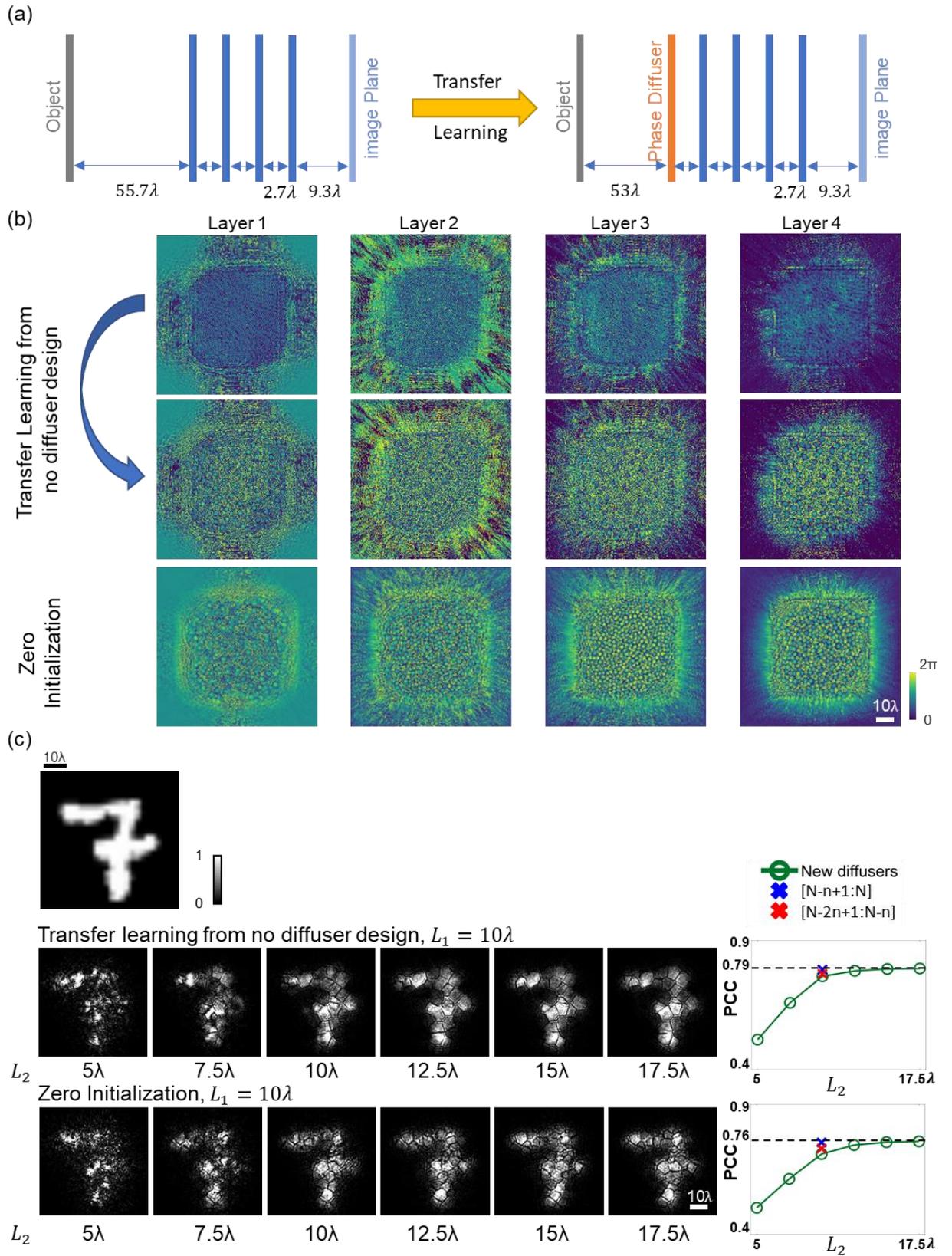

**Figure 7. Impact of diffractive layer initialization.** (a) Schematic of transfer learning from a diffractive design trained without any diffusers. (b) The diffractive layers trained without diffusers (used as initial condition, top row), the diffractive layers after transfer learning (middle row), and the diffractive layers trained with zero-phase initialization (bottom row). The diffractive neural networks that aim to see through random diffusers were trained with $L_1 = 10\lambda$. (c) Visualization and PCC values of the diffractive networks tested under $L_2$: $5\lambda$, $7.5\lambda$, $10\lambda$, $12.5\lambda$, $15\lambda$ and $17.5\lambda$. The contrast of the images was enhanced for visualization (see the Methods section for details).

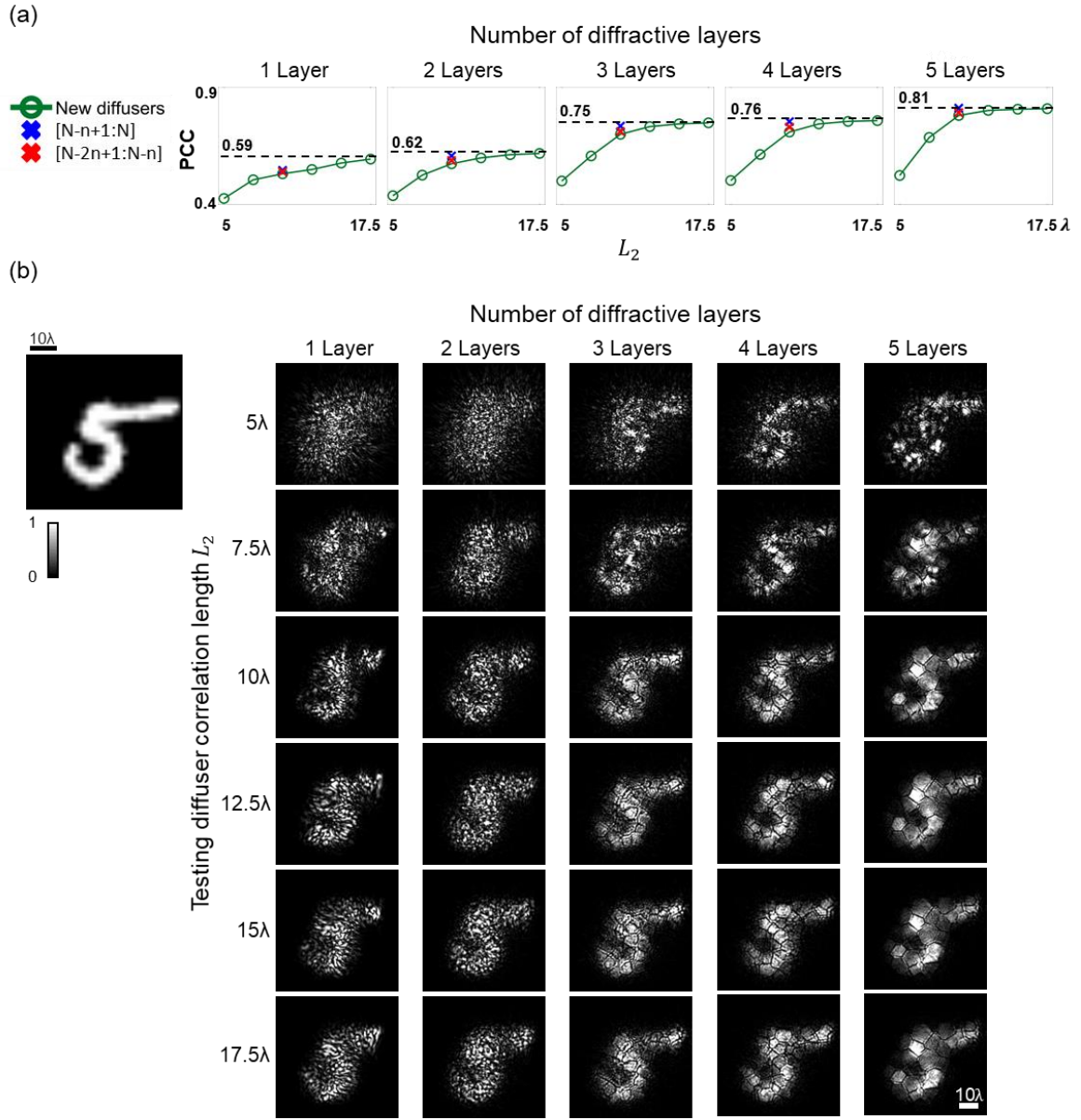

**Figure 8. Depth feature of diffractive optical networks.** (a) PCC values of diffractive neural networks with 1, 2, 3, 4 and 5 layers, trained with random phase diffusers, $L_1 = 10\lambda$. All the resulting diffractive models were tested with new phase diffusers with correlation lengths of $L_2$: $5\lambda$, $7.5\lambda$, $10\lambda$, $12.5\lambda$, $15\lambda$ and $17.5\lambda$. (b) Visualization of the results in (a). The contrast of the images was enhanced for visualization (see the Methods section for details).

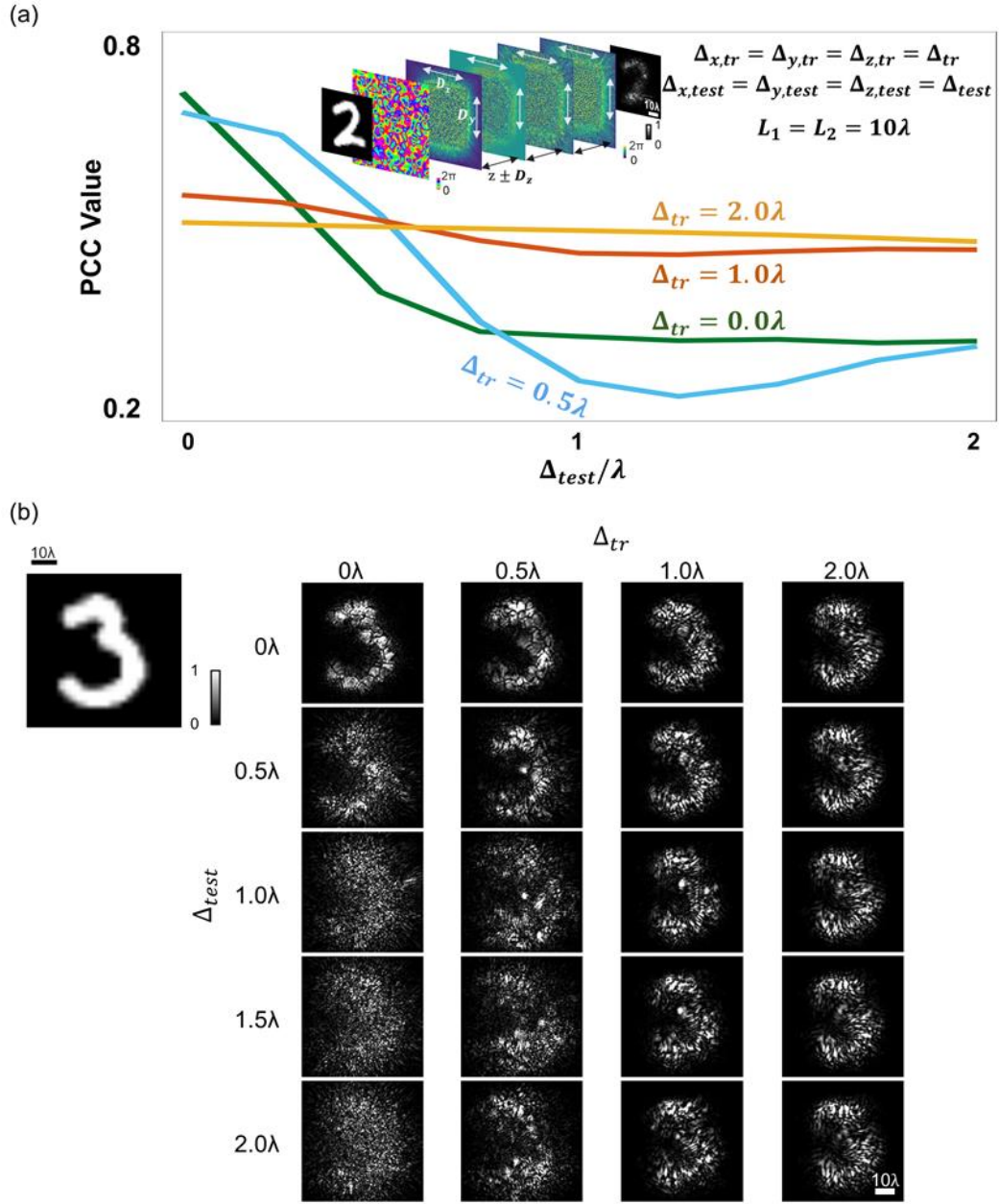

**Figure 9. Vaccinated diffractive neural networks.** (a) The image reconstruction PCC values of diffractive neural networks trained with different levels of vaccination against physical misalignments. Insert: a schematic of the diffractive layer misalignment. (b) Visualization of results in (a) with $\Delta_{test} = 0.0\lambda,\ 0.5\lambda,\ 1.0\lambda,\ 1.5\lambda$ and $2,0\lambda$. The contrast of the images was enhanced for visualization (see the Methods section for details).